\documentclass{article}

\usepackage{PRIMEarxiv}

\usepackage[utf8]{inputenc} 
\usepackage[T1]{fontenc}    
\usepackage{hyperref}       
\usepackage{url}            
\usepackage{booktabs}       
\usepackage{amsfonts}       
\usepackage{nicefrac}       
\usepackage{microtype}      
\usepackage{lipsum}
\usepackage{fancyhdr}       
\usepackage{graphicx}       
\graphicspath{{media/}}     

\usepackage{amsmath,amssymb}
\usepackage{bm}
\title{Self-consistent radiative backaction in dispersion interactions: a minimal mQED model}

\author{
  Johannes Fiedler \\
  Department of Physics and Technology\\
  University of Bergen\\
  All\'egaten 55, 5007 Bergen, Norway\\
  \texttt{johannes.fiedler@uib.no} \\
}

\begin{document}
\maketitle

\begin{abstract}
Dispersion interactions are usually derived assuming fixed internal spectra of the interacting quantum systems. Here, we relax this assumption and study how self-consistent electromagnetic backaction modifies van der Waals interactions when excitation energies and transition dipole moments are allowed to respond to the interaction itself.

Within a macroscopic quantum electrodynamics framework, we formulate a self-consistent treatment that includes both self-energy corrections and mutual backaction. Using a minimal three-level model, we show that, while one-sided self-energy effects are short-ranged, fully self-consistent backaction can lead to substantial, long-ranged modifications of the effective van der Waals interaction.

Our analysis demonstrates that these effects originate from the coherent accumulation of repeated photon-mediated scattering processes. The results highlight limitations of perturbative dispersion theories with fixed spectra and identify few-level systems as a clean platform for studying backaction in dispersion forces.
\end{abstract}


\section{Introduction}
\label{sec:introduction}

Dispersion interactions arising from quantum electromagnetic fluctuations are a universal feature of matter, governing forces and energy shifts across a wide range of length scales. Since their original description by London for non-retarded interactions and their extension to the retarded regime by Casimir and Polder, these forces have been recognised as a direct consequence of the coupling between polarisable quantum systems and the quantised electromagnetic field~\cite{London1930,CasimirPolder1948}. At the molecular scale, dispersion interactions contribute decisively to binding energies, molecular structure, and cohesion, while at larger scales they play a central role in colloidal stability, nanoscale transport, and surface-mediated forces~\cite{IsraelachviliBook,StoneBook,RevModPhys.88.045003,Review2023}.

Within quantum electrodynamics, dispersion forces are commonly formulated in terms of Casimir--Polder (CP) potentials, which describe position-dependent energy shifts induced by vacuum and medium-assisted electromagnetic fluctuations~\cite{PhysRevA.28.2649,PhysRevA.28.2663}. In this framework, the interaction energy is expressed through the electromagnetic Green tensor and the linear response of the interacting particles, typically encoded in their dynamic polarisabilities. This approach has been successfully applied to atoms and molecules in free space, near macroscopic bodies, and in structured photonic environments~\cite{WylieSipe1984,WylieSipe1985,BuhmannScheel2008,BuhmannBook}.

In most practical applications, dispersion interactions are evaluated under the assumption that the internal spectra of the interacting particles remain fixed~\cite{PhysRevLett.130.200401,PhysRevA.91.013614,Galiffi_2017,PhysRevLett.125.050401}. Notable extensions include situations involving excited states and resonant processes, where long-range oscillatory interactions and virtual resonant emission can occur~\cite{PhysRevLett.118.123001,PhysRevA.95.022704}. Radiative self-energy effects, such as Lamb shifts~\cite{PhysRev.72.241}, level broadenings~\cite{D0CP02863K}, or weak modifications of transition dipole moments~\cite{10.1063/5.0106503}, are either neglected or treated as small, local corrections that do not qualitatively affect the interaction~\cite{Milonni1994}. In the present work, we consider environment-induced modifications of the internal response within macroscopic quantum electrodynamics, where such corrections arise from finite electromagnetic Green tensors. In contrast to ultraviolet renormalisation in bound-state QED, no divergences or counterterms appear in this framework, and the resulting self-energy contributions can be interpreted as a medium-dependent dressing of the particle response. As a consequence, dispersion forces are usually computed perturbatively, assuming a clear separation between the internal structure of the particles and their mutual electromagnetic coupling.

This separation becomes increasingly questionable at very small separations, comparable to typical molecular bond lengths or intermolecular distances in condensed phases. In this regime, dispersion interactions are no longer merely weak background forces but compete with chemical bonding, charge redistribution, and environmental screening. Such conditions are encountered, for example, during molecular association and reaction pathways, in catalysis at surfaces or within cavities, and in transport processes through membranes or confined soft-matter systems~\cite{IsraelachviliBook,Persson2005,Zwanzig1990}. Here, even modest radiative modifications of excitation energies or transition strengths may be coherently amplified through repeated photon-mediated interactions.

Macroscopic quantum electrodynamics has shown that electromagnetic self-energy effects can go beyond simple level shifts. In structured or confined environments, the electromagnetic field can induce coherent mixing of internal states, modify transition strengths, and alter effective selection rules~\cite{BuhmannScheel2008,Intravaia2014}. While such effects have been extensively studied for single particles near interfaces or in cavities, their consequences for inter-particle dispersion interactions have received comparatively little attention.

This raises a fundamental question: how are dispersion forces modified when electromagnetic self-energy and state mixing are treated self-consistently for interacting quantum systems? In particular, can mutual radiative backaction between two particles qualitatively alter the magnitude or spatial dependence of van der Waals interactions beyond what is captured by one-sided or perturbative corrections?

In this work, we address this question by formulating dispersion interactions as a self-consistent problem within macroscopic quantum electrodynamics. We explicitly include photon-mediated self-energy corrections, state mixing, and mutual backaction between two particles on equal footing. This allows us to disentangle local dressing effects from genuinely non-local modifications of the interaction energy. Using a minimal few-level model, we show that while one-sided self-energy corrections remain short-ranged, fully self-consistent backaction can lead to substantial and long-ranged modifications of the effective van der Waals interaction.

Our results highlight intrinsic limitations of dispersion theories based on fixed internal spectra and identify few-level systems as a clean theoretical platform for isolating backaction effects. We emphasise that phenomena related to dense or quasi-degenerate spectra, as well as dissipative broadening, require further investigation and lie beyond the scope of the present work. These aspects are expected to become particularly important in realistic molecular and condensed-phase systems and will be addressed in future studies.

\section{Theoretical framework}
\label{sec:theory}

We formulate dispersion interactions between two polarisable particles within macroscopic quantum electrodynamics (mQED), explicitly accounting for electromagnetic self-energy, state mixing, and mutual radiative backaction in a self-consistent manner. As in standard dipolar mQED, the present description is intended as an effective response-based treatment. At very small separations, additional effects such as higher multipoles, wavefunction overlap, or exchange contributions may become relevant and are beyond the scope of the present work.

The formalism developed in this section can be interpreted as a self-consistent response problem. 
In this picture, polarisability is not treated as a fixed input quantity but is modified by photon-mediated interactions encoded in the electromagnetic Green tensor.

Formally, this corresponds to a Dyson-type equation for the response function, in which the self-energy generates corrections that depend on the dressed system itself. The different levels of approximation considered in this work, bare, one-sided, and fully self-consistent, can then be understood as successive truncations or iterations of this self-consistent structure.

\subsection{Casimir--Polder interaction in macroscopic QED}

In macroscopic quantum electrodynamics, dispersion interactions arise as position-dependent energy shifts induced by the coupling of matter to the quantised electromagnetic field~\cite{Review2023}. For a particle at position $\mathbf{r}$, the Casimir--Polder (CP) potential can be expressed in terms of the electromagnetic Green tensor $\mathbf{G}(\mathbf{r},\mathbf{r}',\omega)$ and the particle's dynamic polarisability $\bm{\alpha}(\omega)$~\cite{BuhmannBook}. For two particles $A$ and $B$, the van der Waals interaction energy reads
\begin{equation}
U_{AB}(r) =  -\frac{\hbar\mu_0^2}{2\pi} \int\limits_0^\infty \mathrm{d}\xi\, \xi^4\,
\mathrm{Tr}\!\left[
\bm{\alpha}_A(\mathrm{i}\xi)\cdot
\mathbf{G}(\mathbf{r}_A,\mathbf{r}_B,\mathrm{i}\xi)\cdot
\bm{\alpha}_B(\mathrm{i}\xi)\cdot
\mathbf{G}(\mathbf{r}_B,\mathbf{r}_A,\mathrm{i}\xi)
\right]\,,
\label{eq:vdw_standard}
\end{equation}
where $r=|\mathbf{r}_A-\mathbf{r}_B|$ denotes the interparticle separation.

In conventional treatments, the polarisabilities $\bm{\alpha}_{A,B}(\omega)$ are taken as fixed molecular response functions, independent of the electromagnetic environment and of the presence of other particles. Equation~\eqref{eq:vdw_standard} then yields the familiar $U(r)\propto -C_6/r^6$ behaviour in the non-retarded regime.

The different approximation levels considered in this work can be understood as successive stages of a self-consistent response problem, which connect naturally to standard situations in dispersion theory. In the conventional ``bare'' approach, the internal spectra of both particles are kept fixed, such that the polarisabilities remain independent of the interparticle separation. This corresponds to the standard treatment of van der Waals interactions between ground-state atoms and molecules, as discussed, e.g., in Ref.~\cite{BuhmannBook}.

A one-sided treatment, in which only one particle is dressed by the electromagnetic environment, corresponds to the well-established situation of a particle in a modified photonic environment, such as near a surface or in a cavity, where radiative self-energy effects lead to shifts of transition frequencies and modifications of dipole matrix elements (e.g., Purcell-type effects~\cite{10.1063/5.0073503,10.1063/5.0106503}). In this case, the response of one particle is modified, while the second particle acts only as a probe and does not feed back onto the first.

The fully self-consistent backaction scheme considered here goes one step further: the response functions of both particles become interdependent through the electromagnetic field, such that modifications of one particle affect the environment of the other and vice versa. This mutual dressing becomes relevant when the dimensionless mixing parameters $c_{km}(r)$, introduced below, are no longer negligible, and it requires a self-consistent treatment of both particles on equal footing. In this sense, the present framework extends the standard use of environment-modified response functions by allowing the environment itself to become dynamical through the presence of other particles.

\subsection{Self-energy and state mixing}

Beyond lowest-order perturbation theory, the interaction with the electromagnetic field induces a self-energy correction to the internal Hamiltonian of each particle~\cite{PhysRevA.60.4094}. In mQED, this correction is described by a frequency-dependent self-energy operator,
\begin{equation}
\bm{\Sigma}(\omega)=\frac{1}{\hbar}\,\mathbf{d}\cdot\mathbf{G}(\mathbf{r},\mathbf{r},\omega)\cdot\mathbf{d}\,,
\end{equation}
where $\mathbf{d}$ denotes the dipole operator of the particle. The self-energy generally contains both diagonal contributions, corresponding to Lamb-shift–type level shifts, and off-diagonal terms that coherently mix internal states~\cite{Ribeiro_2015}. We emphasise that the self-energy introduced here should not be interpreted as a renormalised bound-electron self-energy in the sense of quantum electrodynamics. Within macroscopic QED, the Green tensor is finite and encodes the electromagnetic environment, such that $\bm{\Sigma}(\omega)$ represents a medium-dependent, frequency-resolved dressing of the internal degrees of freedom rather than a quantity requiring ultraviolet regularisation.

In structured or strongly coupled electromagnetic environments, such off-diagonal terms can lead to appreciable state mixing, modifying transition energies and effective dipole moments. Importantly, even when individual self-energy corrections remain small, their impact on the molecular response can be amplified through repeated photon-mediated interactions.

Throughout this work, we restrict attention to discrete, non-degenerate excitation spectra. The treatment of exact or quasi-degenerate levels, as well as continuous excitation continua, can give rise to genuine resonances and non-perturbative behaviour and lies beyond the scope of the present analysis.

\subsection{Environment-induced modification of the dynamic polarisability}

The environment-induced state mixing has direct consequences for the linear electromagnetic response. For an isolated particle in eigenstate $|m\rangle$, the dynamic electric polarisability is given by
\begin{equation}
\alpha^{(0)}_{ij}(\omega) = \frac{1}{\hbar} \sum_k \left( \frac{d_i^{mk} d_j^{km}}{\omega_{km}-\omega} + \frac{d_j^{mk} d_i^{km}}{\omega_{km}+\omega} \right)\,.
\label{eq:alpha_free}
\end{equation}
In general, the polarisability involves contributions from a complete set of discrete and continuum states. In the present work, we restrict our attention to a discrete set of levels, thereby allowing us to isolate the effect of environment-induced state mixing in a controlled manner. Continuum contributions, while quantitatively important for realistic systems, are not required to illustrate the qualitative backaction mechanism discussed here and are therefore omitted for clarity.

When the dressed states are used instead, the polarisability acquires additional contributions. To leading order in the self-energy, it can be written as
\begin{equation}
\bm\alpha(\omega) = \bm\alpha^{(0)}(\omega) + \Delta\bm\alpha(\omega)\,,
\end{equation}
where $\Delta\bm\alpha$ contains two physically distinct contributions and represents the leading-order expansion of the underlying self-consistent response scheme. 
The first term originates from environment-induced state mixing, which modifies the dipole matrix elements,
\begin{align}
\Delta\alpha_{ij}^{(\mathrm{wf})}(\omega) = \frac{1}{\hbar} \sum_{k,c\neq m}
\Bigg[ \frac{ \Sigma_{cm} \, d_i^{ck} d_j^{km} }{ (E_m-E_c)(\omega_{km}-\omega) } + \frac{ \Sigma_{cm} \, d_j^{ck} d_i^{km} }{ (E_m-E_c)(\omega_{km}+\omega) } + \mathrm{c.c.} \Bigg]\,,
\label{eq:delta_alpha_wf_new}
\end{align}
while the second term arises from the shift of the transition frequencies, $\omega_{km}\rightarrow\omega_{km}+\Delta\omega_{km}$,
\begin{equation}
\Delta\alpha_{ij}^{(\mathrm{en})}(\omega) = -\frac{1}{\hbar} \sum_k \left[ \frac{\Delta\omega_{km}}{(\omega_{km}-\omega)^2} d_i^{mk}d_j^{km} + \frac{\Delta\omega_{km}}{(\omega_{km}+\omega)^2} d_j^{mk}d_i^{km} \right]\,.
\end{equation}

The total correction $\Delta\bm\alpha=\Delta\bm\alpha^{(\mathrm{wf})} +\Delta\bm\alpha^{(\mathrm{en})}$ depends explicitly on the electromagnetic Green tensor evaluated at the particle position and therefore on the surrounding environment. In this sense, the procedure corresponds to a self-consistent dressing of the linear response, which effectively resums repeated photon-mediated interactions in analogy to a Dyson equation. We note that this self-consistent dressing of the response bears a structural resemblance to the structure of linear-response time-dependent density functional theory (TDDFT), where the interacting response function is obtained from a Dyson-like equation involving the non-interacting response and an effective interaction kernel~\cite{RungeGross1984,Casida1995,Onida2002}. In the present context, it is not the electronic density but the polarisability that is updated self-consistently through photon-mediated interactions.

\subsection{Mutual radiative backaction}

When two particles are present, each particle's electromagnetic environment is modified by the other. This modification is mediated by the non-local structure of the electromagnetic Green tensor, which, in the presence of a second particle, contains not only the free-space contribution but also the field scattered by that particle. This leads to mutual radiative backaction: the self-energy of particle $A$ depends on the response of particle $B$ and vice versa. In particular, the Green tensor entering the self-energy of particle $A$ acquires a contribution that depends on the polarisability of particle $B$, such that the dressing of each particle becomes explicitly dependent on the state of the other. A consistent description, therefore, requires solving the dressing of both particles self-consistently. This mutual dependence gives rise to a feedback mechanism in which modifications of the internal structure of one particle affect the electromagnetic field experienced by the other, which in turn feeds back onto the first. Radiative backaction thus emerges as a genuinely non-local effect mediated by photon exchange, rather than from purely local self-energy corrections.

We distinguish three levels of approximation:
(i) the conventional bare interaction with fixed response functions,
(ii) a one-sided self-energy correction, where only one particle is dressed by the electromagnetic environment, and
(iii) a fully self-consistent backaction treatment, in which both particles are mutually dressed through photon-mediated coupling.

The fully self-consistent approach corresponds to a Dyson-type resummation of photon-mediated interactions at the level of the response function, thereby summing an infinite sequence of photon-exchange processes between the particles. In this interpretation, the different approximation levels introduced above can be understood as successive truncations of this self-consistent scheme: the conventional treatment with fixed polarisabilities corresponds to the zeroth-order approximation, the one-sided correction to the first iteration, and the fully self-consistent solution to the infinite resummation. While each individual contribution may be perturbatively small, their coherent accumulation can lead to qualitative modifications of the effective dispersion interaction, including changes in magnitude and spatial range.

From a diagrammatic perspective, this corresponds to repeated photon-exchange processes that couple the self-energy corrections of both particles.

\subsection{Scope and limitations}

The framework developed here is designed to isolate the role of electromagnetic self-energy and mutual radiative backaction in dispersion interactions within a minimal and controlled setting. 
We restrict the analysis to discrete excitation spectra and neglect non-radiative decay and many-body effects, focusing on regimes in which environment-induced corrections remain small compared to intrinsic level separations.

These assumptions allow us to disentangle local self-energy dressing from genuinely non-local modifications of dispersion forces arising from photon-mediated coupling. 
Extensions to quasi-degenerate systems or to continuous spectra are expected to introduce additional resonant features, but are not expected to alter the basic mechanism identified here.

\section{Model systems}
\label{sec:models}

To make the general framework explicit, we now analyse a hierarchy of model systems of increasing complexity. 
Rather than aiming at quantitative material-specific predictions, these models are chosen to expose the minimal ingredients required for self-consistent backaction to modify dispersion interactions. In particular, the three-level system represents the minimal configuration in which off-diagonal self-energy terms can induce coherent state mixing and thereby modify the linear response. While more complex spectra distribute this mechanism over many transitions, they are not expected to qualitatively alter its origin.

While the three-level model isolates the minimal mechanism, realistic atomic and molecular systems involve a larger number of discrete and continuum states contributing to the polarisability. In such cases, the same backaction mechanism is expected to be distributed over multiple transitions, leading to a smoother but qualitatively similar modification of the effective interaction. In such cases, the same backaction mechanism is distributed over multiple transitions, leading to a smoother but qualitatively similar modification of the effective interaction.

Throughout this section, we consider particles in free space, described by the vacuum Green tensor. 
This isolates radiative self-energy and mutual backaction effects arising purely from photon-mediated coupling, without additional contributions from boundaries or material environments.

Unless stated otherwise, we focus on parameter regimes in which unperturbed level separations remain large compared to self-energy corrections. This ensures that the resulting state mixing can be analysed within a controlled perturbative framework while retaining the essential backaction mechanism.

\subsection{Free-space Green tensor}
\label{subsec:green}

The electromagnetic response of the vacuum is fully characterised by the free-space dyadic Green tensor $\mathbf G^{(0)}(\mathbf r,\mathbf r',\omega)$, which satisfies
\begin{equation}
\nabla\times\nabla\times \mathbf G^{(0)}(\mathbf r,\mathbf r',\omega)
-\frac{\omega^2}{c^2}\mathbf G^{(0)}(\mathbf r,\mathbf r',\omega)
= \mathbf I\,\delta(\mathbf r-\mathbf r') \,,
\end{equation}
subject to outgoing-wave boundary conditions.
For dispersion-force calculations, it is natural to work on the imaginary frequency axis $\omega=\mathrm i\xi$, where the vacuum Green tensor takes the well-known closed form~\cite{PowerThiru,BuhmannBook}
\begin{equation}
\mathbf G^{(0)}(\mathbf r,\mathbf r',\mathrm i\xi)
= \frac{1}{4\pi r^3}
\left[
(1+x+x^2)\mathbf I
-(3+3x+x^2)\hat{\mathbf r}\hat{\mathbf r}
\right]\mathrm e^{-x}\,,
\qquad x=\frac{\xi r}{c}\,,
\label{eq:G0_imag}
\end{equation}
where $r=|\mathbf r-\mathbf r'|$ and $\hat{\mathbf r}=(\mathbf r-\mathbf r')/r$.

In the non-retarded limit $x\ll1$, Eq.~\eqref{eq:G0_imag} reduces to the familiar near-field form
\begin{equation}
\mathbf G^{(0)}(\mathbf r,\mathbf r',\mathrm i\xi)
\simeq -\frac{c^2}{4\pi\xi^2 r^3}
\left(\mathbf I-3\hat{\mathbf r}\hat{\mathbf r}\right)\,,
\end{equation}
which underlies the conventional $U(r)\propto -C_6/r^6$ scaling of the vdW interaction. At larger separations, retardation suppresses contributions from high imaginary frequencies and introduces a smooth crossover to the Casimir--Polder regime~\cite{CasimirPolder1948}.

The full Green tensor~\eqref{eq:G0_imag} therefore provides a physically well-defined frequency-dependent regularisation of self-energy contributions and plays a central role in the backaction effects discussed below.

\subsection{Three-level system}
\label{subsec:threelevel}

As a minimal microscopic model capturing environment-induced state mixing and its feedback onto dispersion forces, we consider two identical particles separated by a distance $r$, each described by three discrete internal states $\lvert 0\rangle$, $\lvert 1\rangle$, and $\lvert 2\rangle$. The corresponding transition frequencies are denoted by $\omega_{10}$ and $\omega_{20}$ with $\omega_{20}>\omega_{10}$. We emphasise that this model is not intended to represent a specific atomic or molecular spectrum, but rather to isolate the minimal mechanism by which off-diagonal self-energy terms induce coherent state mixing. Rather, it constitutes the minimal model in which off-diagonal self-energy terms can induce coherent mixing of internal states, thereby modifying the particle's linear response. In realistic systems, the polarisability involves a large number of discrete and continuum states. The mechanism identified here is therefore not restricted to a few-level system, but is expected to be distributed over many transitions. The present model isolates this mechanism in a controlled and transparent setting, without additional complexity arising from dense spectra or near-degeneracies. In this sense, the three-level model plays a role analogous to minimal models used in quantum optics, where essential physical mechanisms can be identified independently of the full spectral complexity of real systems.

The internal structure is characterised by electric-dipole transition matrix elements $\mathbf d_{01}$, $\mathbf d_{02}$, and $\mathbf d_{12}$, which we assume to be isotropic for simplicity. This three-level structure constitutes the minimal setting in which off-diagonal self-energy terms can coherently mix excited states via the electromagnetic field, while remaining sufficiently simple to allow for a transparent physical interpretation.

In the numerical examples presented below, we choose representative optical-scale parameters,
\begin{equation}
\hbar\omega_{10}=2.0~\mathrm{eV}\,, \qquad
\hbar\omega_{20}=3.0~\mathrm{eV}\,,
\end{equation}
and transition dipole moments of the order of a few Debye,
\begin{equation}
|\mathbf d_{01}|=3.0~\mathrm{D}\,, \qquad
|\mathbf d_{02}|=2.2~\mathrm{D}\,, \qquad
|\mathbf d_{12}|=1.0~\mathrm{D}\,.
\end{equation}
These values are typical for small molecules and atoms, ensuring that the system remains well within the perturbative regime at nanometre separations.

The environment-induced self-energy modifies both transition frequencies and dipole moments, thereby affecting the dispersion interaction itself. To quantify this effect, we define an effective, distance-dependent van der Waals coefficient
\begin{equation}
C_6^{\mathrm{eff}}(r)=-U(r)\,r^6\,,
\end{equation}
and normalise it to the bare free-space value $C_6$.

\begin{figure}[t]
  \centering
  \includegraphics[width=0.6\columnwidth]{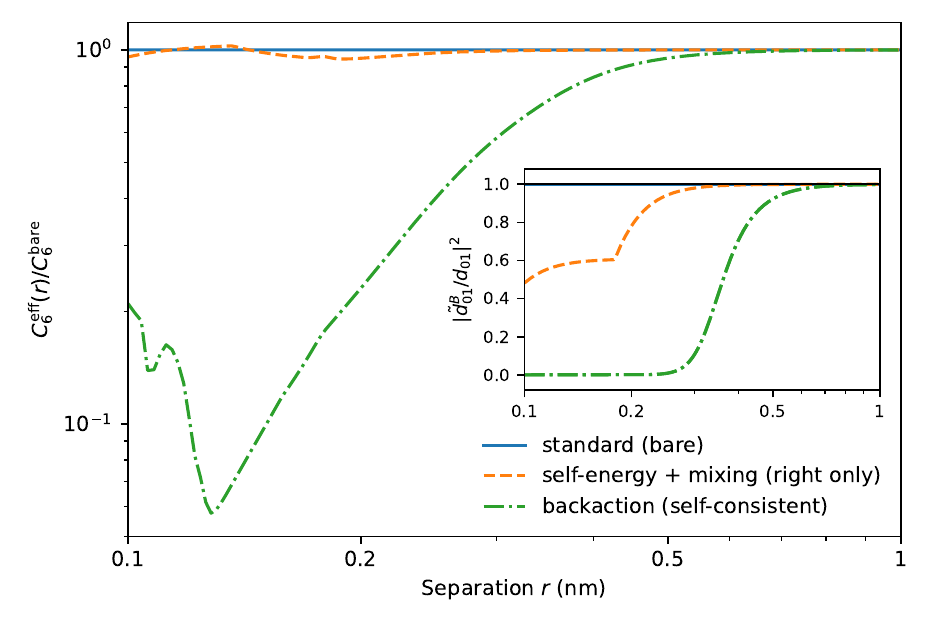}
\caption{Ratio $C_6^{\mathrm{eff}}(r)/C_6$ for the three-level system. Solid line: bare interaction. Dashed line: self-energy and state mixing applied to one particle only. Dash-dotted line: fully self-consistent backaction. Inset: corresponding environment-induced modification of the dominant transition dipole moment, quantified by $|\tilde d_{01}/d_{01}|^2$.}
  \label{fig:vdw_ratio_inset}
\end{figure}

Figure~\ref{fig:vdw_ratio_inset} shows the ratio $C_6^{\mathrm{eff}}(r)/C_6$ for three different levels of approximation: (i) the conventional bare interaction, (ii) a one-sided treatment in which self-energy and state mixing are applied to one particle only, and (iii) a fully self-consistent backaction scheme in which both particles are mutually dressed by photon exchange.

At large separations, all curves collapse onto the conventional van der Waals result, recovering the expected $r^{-6}$ scaling. At shorter distances, however, clear deviations emerge. The one-sided correction remains short-ranged, whereas the fully self-consistent backaction results in a significantly stronger, more extended modification of the interaction.

The inset of Fig.~\ref{fig:vdw_ratio_inset} reveals that the onset of backaction correlates directly with a pronounced environment-induced modification of the transition dipole moment due to state mixing. As the separation decreases, off-diagonal self-energy terms coherently admix excited states, redistributing oscillator strength and modifying the effective polarisability.

The resulting non-monotonic behaviour of $C_6^{\mathrm{eff}}(r)$ reflects a competition between dipole dressing and energy-level shifts. While state mixing tends to reduce effective dipole strengths, modified energy denominators in the polarisability partially compensate this reduction. The inclusion of the full free-space Green tensor ensures that this crossover remains smooth and free of unphysical divergences.

Overall, the three-level model demonstrates that radiative backaction in van der Waals interactions arises from a self-consistent modification (dressing) of internal response functions rather than from a breakdown of perturbation theory. It therefore provides a clean and controlled reference system for understanding backaction effects before addressing more complex atomic or molecular spectra.

\subsection{Scaling analysis}
\label{sec:scaling}

The three-level system introduced in Sec.~\ref{subsec:threelevel} provides a minimal and analytically transparent setting in which the parametric origin of radiative backaction effects can be identified. In particular, it allows us to isolate the dimensionless quantity that controls the onset of environment-induced state mixing and its feedback onto dispersion forces.

\subsubsection*{Non-retarded regime}

The strength of radiative mixing between two internal states $\lvert k\rangle$ and $\lvert m\rangle$ is governed by the ratio between the off-diagonal electromagnetic self-energy and the unperturbed level spacing,
\begin{equation}
c_{km}(r)\;\sim\;\frac{\Sigma_{km}(r)}{E_m-E_k}\,.
\label{eq:ckm_def}
\end{equation}
Within macroscopic quantum electrodynamics, the self-energy $\Sigma_{km}$ arises from virtual photon exchange and is proportional to the dipole--dipole coupling mediated by the Green tensor.

In the non-retarded limit $r\ll c/\omega$, the near-field Green tensor scales as $G^{(0)}\sim r^{-3}$, such that the second-order self-energy behaves parametrically as
\begin{equation}
\Sigma_{km}(r)\sim \frac{d^2}{r^6}\,,
\end{equation}
where $\ell$ labels intermediate states. Introducing a characteristic dipole moment $d$ and excitation energy scale $\Delta E$, one obtains the scaling estimate
\begin{equation}
c_{km}(r)\sim \frac{d^2}{\Delta E\,r^6}\,.
\label{eq:scaling_mixing}
\end{equation}
For typical atomic and molecular parameters, this ratio remains small at nanometre separations, which explains why conventional van der Waals theory—based on fixed, environment-independent polarisabilities—successfully describes most dispersion interactions~\cite{PowerThiru,SalamBook}.

\subsubsection*{Role of retardation and regularisation}

The above estimate suggests a rapid growth of radiative mixing at short distances. However, this apparent divergence is an artefact of the strict non-retarded approximation. Within the present mQED framework, both the dispersion interaction and the backaction-induced self-energy corrections are mediated by the same electromagnetic Green tensor. The issue is therefore not the emergence of a separate short-distance mechanism, but the behaviour of the common Green-tensor kernel when the full frequency dependence is retained.

When the full frequency dependence of the electromagnetic field is retained, the imaginary-frequency Green tensor acquires an exponential damping factor $\exp(-\xi r/c)$,
\begin{equation}
\mathbf G^{(0)}(\mathbf r,\mathbf r',\mathrm i\xi)\propto \mathrm e^{-\xi r/c}\,,
\end{equation}
which suppresses contributions from frequencies $\xi\gtrsim c/r$~\cite{BuhmannBook}.

As a result, the same retardation-induced damping that regularises the dispersion interaction also controls the self-energy contributions entering the backaction problem. The self-energy integrals entering $\Sigma_{km}$ are effectively cut off at $\xi\sim c/r$, leading to a smooth crossover rather than a divergence. The statement ``smooth crossover rather than a divergence'' therefore refers to the behaviour of this effective dipolar Green-tensor description once the full frequency dependence is retained, in contrast to the strict near-field approximation.

Retardation therefore provides a physical regularisation mechanism~\cite{Parsegian2006} that limits the magnitude of state mixing at short distances and ensures that the self-consistent backaction problem remains well behaved.

\subsubsection*{Backaction length scale}

Combining the non-retarded scaling~\eqref{eq:scaling_mixing} with the retardation-induced cutoff, one can identify a characteristic distance scale $r_\ast$ at which radiative mixing becomes appreciable,
\begin{equation}
r_\ast \sim \left(\frac{d^2}{\Delta E}\right)^{1/6}.
\label{eq:rstar}
\end{equation}
For $r\gg r_\ast$, the mixing coefficients $c_{km}$ are parametrically small and backaction effects are negligible. For $r\lesssim r_\ast$, state mixing and environment-induced dipole modification become visible, but the exponential damping in the Green tensor prevents unphysical growth at even shorter separations.

Importantly, this length scale does not represent a distinct interaction mechanism. Rather, it marks the distance at which a self-consistent treatment of internal response functions becomes necessary. The resulting modification of the van der Waals interaction arises from the coherent accumulation of many weak photon-mediated scattering processes, which are automatically resummed in the self-consistent backaction scheme.

\subsubsection*{Implications for dispersion forces}

The scaling analysis clarifies why self-energy corrections applied to a single particle remain short-ranged, while mutual backaction between two particles can lead to more extended modifications of the effective $C_6$ coefficient. Although the individual mixing amplitudes remain small, their repeated and coherent feedback into the polarisability modifies the dispersion interaction over distances that can exceed $r_\ast$ by a significant amount.

The three-level model thus captures, in its simplest form, the physical origin of the distance-dependent $C_6^{\mathrm{eff}}(r)$ observed in Fig.~\ref{fig:vdw_ratio_inset}. It provides a controlled reference point for understanding backaction effects in more complex atomic and molecular systems, while also delineating the limits of validity of the present approach. In particular, situations involving (quasi-)degenerate levels or continuous spectra require an extended treatment beyond the present framework to be described consistently.

\section{Conclusions and outlook}
\label{sec:conclusions}

In this work, we have formulated dispersion interactions as a self-consistent problem within macroscopic quantum electrodynamics. By treating electromagnetic self-energy, state mixing, and mutual radiative backaction on equal footing, we have gone beyond conventional van der Waals theory, which relies on fixed, environment-independent response functions.

Using a minimal three-level model, we demonstrated that radiative backaction modifies dispersion forces not by introducing a new interaction mechanism, but by self-consistently modifying the internal structure of the interacting particles. While one-sided self-energy corrections remain local and short-ranged, mutual backaction leads to a coherent accumulation of photon-mediated scattering processes that can substantially modify the effective van der Waals coefficient over distances that exceed the scale on which local dressing effects become negligible.

A central result of this work is the identification of a characteristic backaction length scale that separates regimes of perturbative and self-consistent behaviour. Retardation plays a crucial dual role in this context: it regularises the short-distance behaviour of radiative self-energy contributions and enables smooth crossovers rather than divergences in the effective interaction. The resulting distance-dependent modification of dispersion forces is therefore a robust consequence of self-consistency, rather than a breakdown of the underlying theory.

The three-level system serves as a controlled reference that captures the essential physics of backaction while avoiding ambiguities associated with dense spectra, quasi-degeneracies, or chemical complexity. In this sense, the present work establishes a clear conceptual framework for understanding how radiative feedback modifies dispersion interactions at small separations.

Several important extensions naturally follow from the present framework. Most notably, systems with continuous or quasi-continuous excitation spectra, as well as near-degenerate internal states, are expected to exhibit enhanced sensitivity to radiative backaction due to resonant amplification of state mixing. A consistent treatment of such situations will require going beyond the discrete-level models considered here, including damping, linewidths, and possibly non-Hermitian effective descriptions.

From a physical perspective, the relevance of backaction effects is expected to be confined to short separations comparable to molecular bond lengths or intermolecular distances in condensed environments. This places the phenomenon in direct connection with processes such as molecular formation, catalysis in confined geometries, and transport across membranes or nanoporous structures, where electromagnetic environments are structured, and particle separations are inherently small.

More broadly, our results highlight that dispersion forces cannot always be regarded as passive background interactions determined solely by fixed molecular properties. Instead, when radiative self-consistency becomes important, dispersion interactions acquire a dynamical character that reflects the mutual dressing of the interacting particles. Exploring this regime systematically, both theoretically and experimentally, remains an open and promising direction for future research.

An interesting and experimentally relevant extension of the present framework concerns dilute gases of light atoms, in particular atomic hydrogen, where dispersion interactions contribute to pressure-dependent frequency shifts in high-precision spectroscopy. In such systems, the interparticle distances can approach the regime where self-consistent modifications of the internal response become non-negligible~\cite{Jentschura_2019}. In the present picture, a distance-dependent effective interaction coefficient $C_6^{\mathrm{eff}}(r)$ would modify the statistical averaging procedures typically used to describe pressure shifts, potentially leading to systematic corrections beyond standard treatments based on fixed polarisabilities. Moreover, in dielectric or structured environments, additional enhancement mechanisms may arise from the combined effects of intrinsic response modifications and the surrounding medium, which alter both the electromagnetic mode structure and the effective coupling between particles. A quantitative analysis of such systems, including realistic atomic spectra and environmental effects, lies beyond the scope of the present study.

\bibliographystyle{unsrt}  
\bibliography{references}

\end{document}